\pdfoutput=1
\documentclass[aps,prd,preprint,superscriptaddress,tightenlines,nofootinbib]{revtex4}

\long\def\symbolfootnote[#1]#2{\begingroup%
\def\thefootnote{\fnsymbol{footnote}}\footnote[#1]{#2}\endgroup}

\usepackage{amsfonts,amsmath,amsthm,amssymb}
\usepackage{mathrsfs}
\usepackage{bm}
\usepackage{color}
\usepackage{epsfig}

\newcommand{\PRE}[1]{{#1}}   

\newcommand{\beq}{\begin{equation}}
\newcommand{\eeq}{\end{equation}}
\newcommand{\bea}{\begin{flushleft} \begin{eqnarray}}
\newcommand{\eea}{\end{eqnarray}\end{flushleft}}

\newcommand{\postscript}[2]{\setlength{\epsfxsize}{#2\hsize}
   \centerline{\epsfbox{#1}}}
\newcommand{\comment}[1]{}

\newcommand{\ci}[1]{}

\newcommand{\lb}{\left(}
\newcommand{\rb}{\right)}

\newcommand{\ba}{\begin{eqnarray}}
\newcommand{\ea}{\end{eqnarray}}
\newcommand{\be}{\begin{equation}}
\newcommand{\ee}{\end{equation}}
\newcommand{\bay}[1]{\left(\begin{array}{#1}}
\newcommand{\eay}{\end{array}\right)}

\def\xk{{\kappa}}

\usepackage[usenames,dvipsnames]{xcolor}
\definecolor{orange}{cmyk}{0,0.5,1,0}
\definecolor{rossoCP3}{cmyk}{0,.88,.77,.40}
\definecolor{graa}{rgb}{0.8,0.8,0.8}
\definecolor{blaa}{rgb}{0.2,0.2,0.6}

\begin{document}

\preprint{
\hfil
\begin{minipage}[t]{3in}
\begin{flushright}
\vspace*{.1in}
MPP--2016--337\\
LMU-ASC 59/16\\
\end{flushright}
\end{minipage}
}

\title{\PRE{\vspace*{0.3in}} \color{rossoCP3}{Minimal left-right
    symmetric intersecting D-brane model
}}
\PRE{\vspace*{0.1in}} 

\author{{\bf Luis A. Anchordoqui}}

\affiliation{Department of Physics and Astronomy,\\  Lehman College, City University of
  New York, NY 10468, USA
\PRE{\vspace*{.05in}}
}

\affiliation{Department of Physics,\\
 Graduate Center, City University
  of New York,  NY 10016, USA
\PRE{\vspace*{.05in}}
}

\affiliation{Department of Astrophysics,\\
 American Museum of Natural History, NY
 10024, USA
\PRE{\vspace*{.05in}}
}

\author{{\bf Ignatios Antoniadis}}
\affiliation{LPTHE, UMR CNRS 7589\\
Sorbonne Universit\'es, UPMC Paris 6, 75005 Paris, France
\PRE{\vspace*{.05in}}}

\affiliation{Albert Einstein Center, Institute for Theoretical Physics\\
University of Bern, Sidlerstrasse 5, CH-3012 Bern, Switzerland
\PRE{\vspace*{.05in}}}

\author{{\bf Haim \nolinebreak Goldberg}}
\affiliation{Department of Physics,\\
Northeastern University, Boston, MA 02115, USA
\PRE{\vspace*{.05in}}
}

\author{{\bf Xing Huang}}
\affiliation{
Department of Physics and Center for Field Theory and Particle
Physics,\\  
Fudan University,
220 Handan Road, 200433 Shanghai, China
\PRE{\vspace*{.05in}}
}

\author{{\bf Dieter L\"ust}}

\affiliation{Max--Planck--Institut f\"ur Physik, \\ 
 Werner--Heisenberg--Institut,
80805 M\"unchen, Germany
\PRE{\vspace*{.05in}}
}

\affiliation{Arnold Sommerfeld Center for Theoretical Physics 
Ludwig-Maximilians-Universit\"at M\"unchen,
80333 M\"unchen, Germany
\PRE{\vspace{.05in}}
}

\author{{\bf Tomasz R. Taylor}}

\affiliation{Department of Physics,\\
 Northeastern University, Boston, MA 02115, USA 
 \PRE{\vspace*{.05in}}
}

\affiliation{
Institute of Theoretical Physics, 
Faculty of Physics,\\ University of Warsaw, 
ul. Pasteura 5, 02-093 Warszawa, Poland
\PRE{\vspace*{.05in}}
}

\PRE{\vspace*{.15in}}

\begin{abstract}\vskip 2mm
  \noindent We investigate left-right symmetric extensions of the
  standard model based on open strings ending on D-branes, with gauge
  bosons due to strings attached to stacks of D-branes and chiral
  matter due to strings stretching between intersecting D-branes. The
  left-handed and right-handed fermions transform as doublets under
  $Sp(1)_L$ and $Sp(1)_R$, and so their masses must be generated by
  the introduction of Higgs fields in a bi-fundamental
  $(\bm{2},\bm{2})$ representation under the two $Sp(1)$ gauge groups.
  For such D-brane configurations the left-right symmetry must be
  broken by Higgs fields in the doublet representation of $Sp(1)_R$
  and therefore Majorana mass terms are suppressed by some higher
  physics scale. The left-handed and right-handed neutrinos pair up to
  form Dirac fermions which control the decay widths of the
  right-handed $W'$ boson to yield comparable branching fractions into
  dilepton and dijets channels. Using the most recent searches at
  LHC13 Run II with 2016 data we constrain the $(g_R, m_{W'})$
  parameter space. Our analysis indicates that independent of the
  coupling strength $g_R$, gauge bosons with masses $m_{W'} \agt
  3.5~{\rm TeV}$ are not ruled out. As the LHC is just beginning to
  probe the TeV-scale, significant room for $W'$ discovery remains.
\end{abstract}

\maketitle

\section{Introduction}

The $SU(3)_C \otimes SU(2)_L \otimes U(1)_Y$ standard model (SM) of
particle physics has recently endured intensive scrutiny, with a
dataset corresponding to an integrated luminosity of $12.9~{\rm
  fb^{-1}}$ of 2016 $pp$ collisions at $\sqrt{s} = 13~{\rm TeV}$, and
it has proven once again to be a remarkable structure that is
consistent with all experimental results by tuning more or less 19
free parameters.  However, the SM is inherently an incomplete theory,
as it does not explain all known fundamental physical phenomena. The
most obvious omission is that it does not provide a unification with
gravity.

Superstring theory is the best candidate for a unified theory of all
interactions. Dirichlet branes on which fundamental open string
boundaries are attached are key objects to realize four-dimensional
chiral gauge theories as low-energy effective models from superstring
theory~\cite{Sagnotti:1987tw,Bianchi:1990yu,Polchinski:1995mt,Polchinski:1996na}.  In general, gauge
couplings are described by the string coupling, the Regge slope
parameter $\alpha' = M_s^{-2}$, and partial volumes of compactified
spaces. The conventional assumption is that $M_s$ is of order (but
below) the Planck scale, $M_{\rm Pl}$. Of particular interest is also
the possibility of realistic superstring models with low mass string
scale, $M_s \sim 10~{\rm TeV}$, and large volume
compactifications~\cite{Antoniadis:1998ig}.

Intersecting D-brane models enclose a collection of building block
guidelines, which can be used to manufacture the SM or something very
close to
it~\cite{Blumenhagen:2000wh,Angelantonj:2000hi,Aldazabal:2000cn,Aldazabal:2000dg,Blumenhagen:2000ea,Ibanez:2001nd,Blumenhagen:2001te,Cvetic:2001tj,Cvetic:2001nr}. Within
these models the basic unit of gauge invariance is a $U(1)$ field, and
hence a stack of $N$ identical D-branes sequentially gives rise to a
$U(N)$ theory with the associated $U(N)$ gauge group.  If there exist
many types of D-brane, the gauge group grows into product form $\prod
U(N_i)$, where $N_i$ specifies the number of D-branes in each
stack. (For $N = 2$, the gauge group can be $Sp(1) \cong SU(2)$ rather
than $U(2)$.\footnote{In the presence of orientifolds, one also
  obtains orthogonal and symplectic gauge groups.}) Gauge bosons (and
associated gauginos in a supersymmetric model) arise from strings
terminating on {\em one} stack of D-branes, whereas chiral matter
fermions are realized as the Ramond-sector of open strings stretching
between {\em two} stacks.  For further details, see
e.g.~\cite{Blumenhagen:2005mu,Blumenhagen:2006ci}.

The minimal embedding of the SM particle spectrum requires at least
three brane stacks~\cite{Antoniadis:2000ena} leading to three distinct
models of the type $U(3) \otimes U(2)\otimes U(1)$ that were
classified in~\cite{Antoniadis:2000ena, Antoniadis:2004dt}. Only one
of them (model C of~\cite{Antoniadis:2004dt}) has baryon number as a
gauge symmetry that guarantees proton stability (in perturbation
theory), and can be used in the framework of low mass scale string
compactifications. Besides, since the charge associated to the $U(1)$
of $U(2)$ does not participate in the hypercharge combination, $U(2)$
can be replaced by the symplectic $Sp(1)$ representation of
Weinberg-Salam $SU(2)_L$, leading to a model with one extra $U(1)$
added to the hypercharge~\cite{Berenstein:2006pk}. The SM embedding in
four D-brane stacks leads to many more models that have been
classified in~\cite{Antoniadis:2002qm,Anastasopoulos:2006da}. 
Several detailed and extensive phenomenological analyses have allowed us 
 to blueprint new-physics signals of all these embeddings  that could
 potentially be revealed at the
 LHC~\cite{Anchordoqui:2007da,Anchordoqui:2008ac,Anchordoqui:2008hi,Lust:2008qc,Anchordoqui:2008di,Anchordoqui:2009mm,Anchordoqui:2009ja,Anchordoqui:2011ag,Anchordoqui:2011eg,Anchordoqui:2012wt,Anchordoqui:2014wha,Anchordoqui:2015uea,Anchordoqui:2015jxc,Anchordoqui:2016rve},
in future $e^+e^-$ and $\gamma \gamma$ colliders~\cite{Anchordoqui:2010zs}, as well as in astrophysical~\cite{Anchordoqui:2009bn} and cosmological observations~\cite{Anchordoqui:2012wt,Anchordoqui:2012qu,Anchordoqui:2011nh}.

Curiously, the SM is chiral; it is of considerable interest to
investigate ways to restore the left-right symmetry. In this paper we
examine this possibility within the context of D-brane string
compactifications.  The layout of the paper is as follows. We begin in
Sec.~\ref{s2} with an outline of the basic setting of the minimal
left-right symmetric intersecting D-brane model and a discussion on
similarities and differences of the string inspired gauge structure
and the canonical gauge sector generally used to restore the
left-right
symmetry~\cite{Pati:1974yy,Mohapatra:1974hk,Senjanovic:1975rk,Senjanovic:1978ev,Mohapatra:1979ia}.
Aspects of coupling unification are briefly discussed in
Sec.~\ref{guni}.  After that, in Sec.~\ref{s3} we examine the
constraints placed on the parameter space of the D-brane set up by a
broad range of LHC searches, taking into account the full set of
relevant production and decay channels. Finally, we summarize our
results and draw our conclusions in Sec.~\ref{s4}.

\section{Left-Right Symmetry from Intersecting D-branes}
\label{s2}

The minimal left-right symmetric intersecting D-brane model is
described by the gauge group $U(3)_C \otimes Sp(1)_L \otimes U(1)_L
\otimes Sp(1)_R$~\cite{Blumenhagen:2000ea,Blumenhagen:2015fqn}.  The
D-brane content is given in Table~\ref{table-i}; the mirror branes
$a^*,b^*,c^*,d^*$ are not shown.  The right-handed quarks live at the
intersection of the $Sp(1)_R$ stack of the D-branes and the color
stack.  The right-handed leptons stretch between $Sp(1)_R$ and the
lepton brane $U(1)_L$. The left handed fermions bridge a mirror copy
at the intersections of $U(3)_C \leftrightharpoons Sp(1)_L$ and
$Sp(1)_L \leftrightharpoons U(1)_L$.  The model requires two types of
Higgs to generate the masses for the fermions and gauge bosons.  The
massless spectrum in the $a,b,c,d$ basis is summarized in
Table~\ref{table-ii}. The baryon and lepton number, $B$ and $L$,
respectively; can be identified with the charges
\begin{equation} 
B=
\frac{1}{3} Q_a \quad \quad {\rm and} \quad \quad L=Q_{c}
\end{equation}
of the abelian gauge
factors 
\begin{equation}
U(1)_B = \frac{1}{3} U(1)_a \subset U(3)_C  \quad \quad {\rm and} \quad \quad U(1)_L = U(1)_d \, .
\end{equation}
Note that both $B$ and $L$ are anomalous, but the combination $B-L$ is anomaly free. 

Now a point worth noting at this juncture is that, in principle, one
would like to introduce the left-right symmetry imposing a discrete
$\mathbb{Z}_2$ invariance. This would be automatic if the gauge group
could be $O(4)$, but then open strings would have no spinors and thus
no doublets (under one of the $Sp(1)$) to break the symmetry. One
should therefore impose a $\mathbb{Z}_2$ symmetry upon the exchange of
the $Sp(1)_L$ and $Sp(1)_R$ branes, ${\mathscr P}: Sp(1)_L
\leftrightharpoons Sp(1)_R$. This parity symmetry implies in particular equality of
the two $Sp(1)$ gauge couplings and provides an interesting constraint
for determining all couplings at the string scale. To keep our
discussion general, in what follows we also consider models in which
${\mathscr P}$ is broken explicitly in this Higgs sector. These $\,
\backslash \! \!\! \! \!  {\mathscr P}$ models are peculiarly 
interesting as they could provide gauge coupling unification~\cite{Blumenhagen:2015fqn}.

The first type of Higgs is in a
bi-fundamental $(\bm{2},\bm{2})$ representation under the two
$Sp(1)$. The D-brane configuration may contain one or more  bi-doublet
scalars,
\begin{equation}
\Phi_i = \left(\begin{array}{cc} 
{\phi_{1i}^0} & {\phi_{2i}^+} \\
{\phi_{1i}^-} & {\phi_{2i}^0} \\
\end{array}
\right) , \quad \quad {\rm with} \quad \quad \tilde \Phi_i = \tau_2
\Phi^*_i \tau_2 \,,
\label{phi1}
\end{equation}
to which correspond the covariant derivative
\begin{equation}
{\cal D}^\mu \Phi_i = \partial \Phi_i - i g_L \, \vec W_L^\mu  \cdot \frac{\vec
  \tau}{2} \, \Phi_i + i g_R \,
\Phi_i \, \frac{\vec \tau}{2} \cdot \vec W^\mu_R \,,
\end{equation}
where $g_L$ and $g_R$ are gauge coupling constants, $\vec W^\mu_{L,R}$
are the $Sp(1)_{L,R}$ gauge fields,  and $\vec \tau$ are the Pauli
matrices. The bi-doublet scalars couple to  fermion bilinear $\bar Q_L Q_R$ and $\bar L_L
L_R$, and generate masses for quarks and leptons after spontaneous
symmetry breaking by their vacuum expectation values (VEVs)
\begin{equation}
\langle \Phi_i
\rangle = \frac{1}{\sqrt{2}} \ {\rm diag}(k_{1i},  k_{2i})
\, .
\end{equation}
The left-right symmetry is broken by Higgs fields in the doublet
representation of $Sp(1)_R$ with $B-L =1$.  Note that we are forced to
introduce these Higgs fields in vector-like pairs $H^{u,d}_R$ for
anomaly cancellation.  Here $\langle H^u_R \rangle = \left(^0_{v_1}
\right)$, $\langle H^d_R \rangle = \left(^{v_2}_0\right)$, $v_R =
\sqrt{v_1^2 + v_2^2} \gg k_{1i}, k_{2i}$ is ${\cal O}({\rm TeV})$, and
$\tan \varkappa \equiv v_1/v_2$. The associated left-handed Higgs doublets
$H^u_L$ and $H^d_L$ must also be present to maintain the $\mathbb Z_2$
symmetry. Likewise, $\langle H^u_L \rangle = \left(^0_{v_3}
\right)$, $\langle H^d_L \rangle = \left(^{v_4}_0\right)$, $v_L =
\sqrt{{v_3}^2 + {v_4}^2}$, and  $\tan \beta \equiv v_3/v_4$.
In the analysis below we greatly simplify the discussion by
assuming that $H^u_L$ and $H^d_L$ acquire vanishing,
phenomenologically irrelevant, VEVs which are set to zero throughout;
that is $v_L = 0$.

\begin{table}
\caption{D-brane content  of $U(3)_C \otimes Sp(1)_L  \otimes U(1)_L \otimes Sp(1)_R$.}
\begin{center}
\begin{tabular}{cccc}
\hline 
\hline
~~~~Label~~~~ & ~~~~Stack~~~~  & ~~~~Number of Branes~~~~ & ~~~~Gauge Group~~~~ \\
\hline 
$a $ & Color & $N_a =3$ & ~~~$U(3)_C = SU(3)_C \times U(1)_a$~~~ \\
$b$ & Left & $N_b =1$ & $Sp(1)_L\cong SU(2)_L$ \\
$c$ & Lepton & $N_c =1$ & $U(1)_L$ \\
$d$ & Right & $N_d =1$ & $Sp(1)_{R} \cong SU(2)_R $ \\
\hline
\hline
\end{tabular}
\end{center}
\label{table-i}
\end{table}
\begin{table}
  \caption{Massless left-handed spectrum of $U(3)_C \otimes Sp(1)_L  \otimes U(1)_L \otimes Sp(1)_R$.}
\begin{center}
\begin{tabular}{cccccccc}
\hline
\hline
 ~~~Number~~~ & ~~Fields~~ & ~~~Sector~~~  &
~~~Representation~~~ & ~~~$Q_a$~~~ & ~~~$Q_c$~~~  & ~~~$B-L$~~~ \\
\hline
3 & $Q_L$ & $a \leftrightharpoons b$ &  $(\bm{3},\bm{2},\bm{1})$ & $\phantom{-}1$
& $\phantom{-}0$ &
$\phantom{-}1/3$ \\[1mm] 
3 &  $(Q_R)^c$ &   $a \leftrightharpoons d$ &  $(\bm{\bar
    {3}},\bm{1},\bm{2})$&$-1$ & $\phantom{-}0$ & $-1/3$  \\[1mm]
3 & $L_L$ & $b \leftrightharpoons c$ &  $(\bm{1},\bm{2},\bm{1})$ & $\phantom{-}0$
& $\phantom{-}1$
& $-1$ \\[1mm] 
3 &  $(L_R)^c$  &  $c \leftrightharpoons d$  &   $(\bm{1},\bm{1},\bm{2})$
  & $\phantom{-}0$ & $-1$ & $\phantom{-}1$\\ [1mm] 
\hline
$N_\Phi$ & $\Phi$ & $b \leftrightharpoons d$ & $(\bm{1},\bm{2},\bm{2})$ & $\phantom{-}0$ & $\phantom{-}0$ & $\phantom{-}0$ \\[1mm] 
$N_{H_R}$ & $H^u_R$ & $c \leftrightharpoons d$ & $(\bm{1},\bm{1},\bm{2})$
& $\phantom{-}0$ & $\phantom{-}1$ & $-1$ \\[1mm] 
$N_{H_R}$ & $H^d_R$ & $c \leftrightharpoons d$ & $(\bm{1},\bm{1},\bm{2})$
& $\phantom{-}0$ & $-1$ & $\phantom{-}1$ \\[1mm] 
 $N_{H_L}$ & $H^u_L$ & $b \leftrightharpoons c$ & $(\bm{1},\bm{2},\bm{1})$
& $\phantom{-}0$ & $-1$ & $\phantom{-}1$  \\[1mm] 
$N_{H_L}$ & $H^d_L$ & $b \leftrightharpoons c$ & $(\bm{1},\bm{2},\bm{1})$
& $\phantom{-}0$ & $\phantom{-}1$ & $-1$ \\[1mm] 
\hline
\hline
\label{table-ii}
\end{tabular}
\end{center}
\end{table}

In the quark-mass basis the right-handed charged-current (CC) interaction of
the $W_R^\pm$ boson for quarks is given by
\begin{equation}
\mathscr{L}_{\rm CC} = \frac{g_L}{\sqrt{2}}\overline U_L \gamma^\mu \mathbb{V}_L D_L W_{L\mu}^+ + \frac{g_R}{\sqrt{2}}\overline U_R \gamma^\mu \mathbb{V}_R D_R W_{R\mu}^+ +\text{h.c.}\,,
\end{equation}
where $\mathbb{V}_{L}$ is the SM CKM matrix in the canonical form,  and $\mathbb{V}_R$ its
right-handed equivalent, which has in principle different angles and
five extra phases.  The ${\mathscr P}$-symmetric case forces
$|\mathbb{V}_L| = |\mathbb{V}_R|$, thus avoiding
flavor-changing neutral currents. Since additional parameter freedom
is currently superfluous, in all our calculations we assume the validity of
the relation between the left- and right-handed CKM matrices imposed
by the $\mathbb Z_2$ symmetry. More unconventional
models have been discussed in  e.g.~\cite{Kiers:2002cz,Zhang:2007da,Maiezza:2010ic,Dekens:2014ina}.  It is noteworthy that the
gauge fields $W_{L,R}^\pm$ are not quite mass eigenstates. The two
charged gauge-bosons mix because both of them couple to the bi-doublet
that is charged under the two $Sp(1)$ groups. The mass terms for the
charged gauge-bosons are given by
\begin{equation}
\mathscr{L}_{m_W m_{W'}} = \bigg(W^-_{L\mu} \quad W^-_{R\mu}
\bigg) \ \left( \begin{array}{cc}
                         m_W^2 & \zeta m_W^2  \\
                         \zeta m_W^2 & m_{W'}^2 
                         \end{array}\right)  \
\left(\begin{array}{c} W^{+\mu}_L\\ W^{+\mu}_R \end{array} \right) ,
\end{equation}
where
\begin{equation}
m_W^2= \frac{g_L^2}{4}  \sum_i (k_{1i}^2+k_{2i}^2) \quad \quad {\rm and} \quad \quad
m_{W'}^2= \frac{g_R^2}{4} \ \left[2v_R^2 +\sum_i (k_{1i}^2+k_{2i}^2) \right] \, ,
\end{equation}
and where the mixing is parameterized by an $\mathcal{O}(1)$
coefficient $\xk = g_R/g_L$, with  
\begin{equation}
\zeta = \kappa~ \frac{2\sum_i k_{1i}k_{2i}}{\sum_i (k_{1i}^2+k_{2i}^2)}\,.          
\end{equation}
The mass eigenstates $W_{1,2}$ (where $W_1 = W$ is 
identified as the well-known lighter state and $W_2 = W'$),  are related to the gauge
eigenstates $W_L, W_R$ by a rotation of mixing angle $\phi$, given by 
\begin{equation}
\left(\begin{array}{c} W^{+\mu}_L\\ W^{+\mu}_R \end{array} \right) =
\left(\begin{array}{cc} \cos \phi &-\sin\phi \\ \sin\phi
    &\phantom{-}\cos\phi \end{array} \right) \left( \begin{array}{c}
    W^{+\mu}_1\\ W^{+\mu}_2 \end{array} \right)  \quad \quad {\rm with}
\quad \quad \tan 2\phi ={\frac{-2\zeta m_W^2}{m_{W'}^2-m_W^2}} \ll 1\, .
\end{equation}

It is evident that in D-brane string compactifications the left-right
symmetry cannot be broken by a Higgs field in the triplet
representation of $Sp(1)_R$ with $B-L = \pm 2$, because the open
string of such a massless mode would require four instead of two
ends. As a consequence, there is no equivalent of the seesaw mechanism
to generate the Weinberg term~\cite{Weinberg:1980bf} which gives rise
to Majorana neutrinos. To generate Majorana masses (by $Q_c$ charge
conservation) one needs dim-5 operators (such as $(L_R H_R^u)^2$) which are
expected to be suppressed by the string scale. At the renormalizable
level one can only write a coupling of leptons with the Higgs
bi-doublet, which gives a Dirac mass to the neutrinos. As we show in
the next section, the resulting neutrino mass constrains the $W'$
decay channels, narrowing the parameter space for LHC searches.

The {\it physical} neutral gauge bosons $Z_\mu$, ${Z^\prime}_\mu$ and the
photon $A_\mu$ are related to the weak $Sp(1)_{L,R}$ and $U(1)_{B-L}$ states
${W^3_R}_\mu$, ${W^3_R}_\mu$ and $B_\mu$ by an analogous mixing
matrix. Using $v_R^2 \gg k_{1i}, k_{2i}$ the mass ratio of $W'$ and
$Z'$ is found to be
\begin{equation}
\frac{m_{Z'}^2}{m_{W'}^2} \simeq \frac{\kappa^2 ( 1 - \sin^2
  \theta_W)}{\kappa^2 ( 1 - \sin^2 \theta_W) - \sin^2 \theta_W } > 1 \,, 
\end{equation}
with $\theta_W$ the Weinberg angle and $m_Z \simeq m_W/\cos
\theta_W$~\cite{Brehmer:2015cia,Aguilar-Saavedra:2015iew}.

In the foreground the chiral multiplets harbor a $[U(1)_aSp(1)^2_L]$
mixed anomaly through triangle diagrams with fermions running in the
loop. It is straightforward to see that the only anomaly free
combination is $B-L$. The anomaly of the orthogonal combination is
cancelled by the generalized Green-Schwarz mechanism, wherein closed
string couplings yield classical gauge-variant terms whose gauge
variation cancels the anomalous triangle
diagrams~\cite{Green:1984sg,Witten:1984dg,Dine:1987bq,Lerche:1987qk,Ibanez:1999it}. The
extra abelian gauge field becomes massive by the Green-Schwarz anomaly
cancellation, behaving at low energies as a $Z''$ with a St\"uckelberg
mass in general lower than the string scale by an order of magnitude,
corresponding to a loop factor. Higgs VEVs will also generate
additional mass terms for $Z''$, introducing also some small mixing
with other gauge bosons, of order $({\rm TeV}/M_s)^2$. Note that for
models with low mass string scale, the discovery of $Z''$ is within
the LHC reach.

It is worth commenting on an aspect of this study which may seem
discrepant at first blush.  In principle, Euclidean brane instantons
might contribute to Majorana
masses~\cite{Blumenhagen:2006xt,Ibanez:2006da}.  However, this would
only be possible if the left-right breaking scale is of the order of
the string scale, and consequently the $W'$ and $Z'$ gauge bosons
would be out of the LHC reach. Hence, for a TeV-scale left-right
symmetry breaking, we can generically argue as before that the strong
suppression of Majorana mass terms in D-brane models constrains the
$W'$ decay channels, narrowing the parameter space for LHC searches.

\section{Gauge Coupling Unification}
\label{guni}

For conventional models with $10~{\rm TeV} \ll M_s \alt M_{\rm Pl}$,
there are still theoretical differences among the various frameworks
which unfortunately do not easy lend themselves to experimental
observation. In particular, it would be interesting to see what kind
of unification scales (if any) are predicted by D-brane models in
terms of running in the gauge sector. 

To this end we
can define $\alpha_i = g_i^2/(4\pi)$ as usual, and evolve $\alpha_i^{-1}$ which
becomes simply $d \alpha_i^{-1}/dt = - b_i/2\pi$. The corresponding
$\beta$-coefficients for the SM are found to be
\begin{equation}
(b_s^{\rm SM} ,\ b_L^{\rm SM},\ b_Y^{\rm SM}) = (-7,\ -19/6,\ 41/10) \,,
\end{equation}
and those of the left-right symmetric extension are given by
\begin{equation}
(b_s,\  b_L,\ 
b_R,\  b_{B-L}) = (-7,\ -3,\ -3,\ 4) + (\delta b_s,\ \delta b_L,\ \delta b_R,\
\delta b_{B-L}), 
\end{equation}
where $\delta b_i$ stands for the contributions from additional fields,
not accounted for in the SM. For illustration, the coefficients for
the groups $Sp(1)_L \otimes Sp(1)_R$ include the contribution from one
bi-doublet field $\Phi_{\bm{1}, \bm{2}, \bm{2},0}$. We have
included this field in the $b_i$ directly, since the SM Higgs is $
\Phi_{\bm{1},\bm{2},1/2} \in \Phi_{\bm{1}, \bm{2}, \bm{2},0}$ in our
construction. At the left-right breaking scale $m_{W'}$ the
hypercharge coupling splits into the $SU(2)_R$ and the $U(1)_{B-L}$
coupling according to
\begin{equation}
\frac{1}{\alpha_Y (m_{W'})} = \frac{3}{5}\, \frac{1}{\alpha_R (m_{W'})} + \frac{2}{5}
 \, \frac{1}{ \alpha_{B-L} (m_{W'})}
\, . \label{lambdayR}
\end{equation}
We can use (\ref{lambdayR}) to reduce the system of equations by
eliminating one of the four running gauge couplings, because the orthogonal
combination $-2 \alpha_R^{-1}(m_{W'})/5 + 3 \alpha^{-1}_{B-L}
(m_{W'})/5$ is a free parameter. Note that by gauge couplings we mean the
independent parameters at the string scale, since $U(3)$ unifies the
abelian $g'_s$ with the non-abelian $g_s$  with the
appropriate normalisation: $g'_s (M_s) = g_s (M_s)
/\sqrt{6}$~\cite{Antoniadis:2000ena}. The abelian couplings of the
D-brane model are furthered constrained by the orthogonality
condition~\cite{Anchordoqui:2011eg}. Finding a model which unifies correctly, then simply
amounts to calculating a set of consistency conditions on the $\delta
b_i$.

It is clear that by demanding $\mathbb Z_2$ symmetry we have $\delta
b_R = \delta b_L$ and so, if we impose gauge coupling unification,
(\ref{lambdayR}) gives a left-right breaking scale out of the LHC
reach. On the other hand, if $\delta b_R
\neq \delta b_L$ then $m_{W'}$ could be $\cal O({\rm TeV})$. In addition,
the restoration
of the left-right symmetry can accelerate the running of the
gauge couplings, yielding unification at a relatively low energy of about $
10^{15}~{\rm
  GeV}$, see e.g.~\cite{Arbelaez:2013nga}. Note that for the
canonical left-right symmetric models, Super-K lower limits on
half-life for proton
decay~\cite{Nishino:2012bnw,Abe:2013lua,Takhistov:2015fao} shift the
scale of gauge coupling unification to higher
energies~\cite{Dev:2015pga}. However, this constraint does not apply
to (left-right symmetric) intersecting D-brane models, which have
baryon number as gauge symmetry that guarantees proton stability (in
perturbation theory).  

It is important to stress that for intersecting D-brane constructions, one can just
require $g_L(M_s) = g_R(M_s)$ because of the $\mathbb Z_2$ symmetry
(which makes the group $SO(4)$ effectively). The other gauge couplings
are independent since they correspond to different brane stacks. This
allows for an alternative to gauge coupling unification, with $m_{W'}$
of order TeV.

The role played by the renormalization-group flow in supersymmetric
extensions of left-right symmetric D-brane models has been 
studied in~\cite{Blumenhagen:2015fqn,Blumenhagen:2003jy}.  For $\, \backslash \! \!\! \! \!
{\mathscr P}$ models, it is possible to get gauge coupling unification
together with $m_{W'} \sim {\rm TeV}$. To accommodate the
non-observation of supersymmetry signals with a TeV-scale left-right breaking
we assume the following hierarchy of mass scales
$m_{W'} < M_{\rm SUSY} < M_s$, where $M_{\rm SUSY}$ is the scale
of supersymmetry breaking. For $N_{H_L} = 0$, this sets a lower bound on the number of
vector-like pairs $N_{H_R}$ for given a bi-doublet
configuration $N_\Phi$. For example, $N_\Phi = 1$ requires $N_{H_R} \geq
3$; otherwise $M_{\rm SUSY} < m_{W'}$ that is in conflict with our assumption.  Following the extended survival hypothesis~\cite{delAguila:1980qag}, in the energy regime
$m_{W'} < E < M_{\rm SUSY}$ we take the minimal particle content
of the non-supersymmetric left-right symmetric SM,  that is one scalar
Higgs bi-doublet $\Phi$ and one scalar Higgs doublet $H_R^u$, yielding
\begin{equation}
\left(\delta b_s,\  \delta b_L,\ \delta b_R,\
\delta b_{B-L} \right) = \left(0,\ 0,\  1/6,\ 1/4 \right)\, .
\end{equation} 
The $\beta$-function coefficients in the supersymmetric region are
found to be
\begin{equation}
\left(\delta b_s,\ \delta b_L,\ \delta b_R,\
\delta b_{B-L} \right) = \left(4,\ 3 + N_\Phi,\ 3 + N_{\Phi} + N_{H_R} ,\ 2 + 3
N_{H_R}/2 \right) \, .
\end{equation}
For $m_{W'} \sim 2~{\rm TeV}$, the minimal possible scale
of supersymmetry breaking comes out fairly universal $M_{\rm SUSY}
\sim 19~{\rm TeV}$; conjointly, the $Sp(1)_R$ gauge coupling  does
only vary slightly in the region  $0.48 <
g_R({m_W'}) <0.60$~\cite{Blumenhagen:2015fqn}.

Note that non-minimal D-brane constructs with $10~{\rm TeV} \ll M_s
\alt M_{\rm Pl}$ could have more than one linear combination of
anomalous $U(1)$ that are non-anomalous. Under certain topological
conditions the associated gauge bosons can remain massless and obtain
a low mass scale via the ordinary Higgs
mechanism~\cite{Cvetic:2011iq}. Interestingly, the extended scalar
sector of these setups can be used to stabilize the vacuum up to the
string scale~\cite{Anchordoqui:2012fq}.  Some phenomenological aspects
of these kind of $U(1)$ gauge bosons and the prospects to search for
them at the LHC were analyzed elsewhere~\cite{Anchordoqui:2012wt}.

As a matter of fact, the LHC8 phenomenology and discovery reach of massive $Z'$ and $Z''$
gauge bosons have been discussed in detail in our previous
publications~\cite{Anchordoqui:2011ag,Anchordoqui:2011eg,Anchordoqui:2012wt}. The
constraints on the parameter space are largely dominated by dijet
searches. Since the dijet limits from LHC8 with an integrated
luminosity of about $20~{\rm
  fb}^{-1}$~\cite{Chatrchyan:2013qha,Aad:2014aqa} are comparale to
those of LHC13 with above about $12~{\rm
  fb}^{-1}$~\cite{Khachatryan:2015dcf,ATLAS:2016lvi,CMS:2016wpz}, in
the next section we focus attention on $W'$ searches.

\section{LHC Phenomenology: Constraints from $\bm{W'}$ searches}
\label{s3}

The ATLAS and CMS experiments are actively looking for $W'$ and $Z'$
gauge bosons. In particular, searches for $W'$ resonances have been
carried out at LHC8 and LHC13 considering a sequential SM
$W'$~\cite{Altarelli:1989ff} and the usual decay modes: the leptonic
channels ($W' \to \tau \nu$~\cite{Khachatryan:2015pua,CMS:2016ppa},
$W' \to \mu \nu$~\cite{Chatrchyan:2011dx}, $W' \to e
\nu$~\cite{Khachatryan:2010fa}, $W' \to
l\nu$~\cite{Khachatryan:2014tva,Aad:2012dm,Chatrchyan:2012meb,CMS:2015kjy}),
the $W' \to WZ$
channel~\cite{Aad:2015owa,Khachatryan:2014gha,Chatrchyan:2012kk,Khachatryan:2014xja,Aad:2013pdy},
the
dijet~\cite{Chatrchyan:2013qha,Aad:2014aqa,Khachatryan:2015dcf,ATLAS:2016lvi,CMS:2016wpz}
and the $tb$
modes~\cite{Khachatryan:2015edz,Chatrchyan:2014koa,Chatrchyan:2012gqa,Aad:2014xra,Aad:2014xea,CMS:2016ude}.
The sequential SM contains extra heavy neutral bosons $Z'$ and $W'$,
with the same couplings to fermions and bosons as the $Z$ and $W$. The
limits derived by LHC experiments then apply directly to $\mathscr P$
models with $\mathbb Z_2$ symmetry. The lower bound on the mass of
$W'$ is $3.5~{\rm TeV}$ at 95\% C.L~\cite{ATLAS:2016lvi}. In this
section we will extract from the results of the LHC analyses 95\%
C.L. exclusion regions on the $(m_{W'} , g_R)$ plane for $\,
\backslash \! \!\! \! \!  {\mathscr P}$ models, in which $g_R <
g_L$.

The decay rate of $W'{}^+$ to fermion pairs is found to be
\begin{align}
  \label{eq:BRjj}
  \Gamma(W'{}^+ \to u \bar{d}) &= \Gamma(W'{}^- \to e^- \nu) = 3 \kappa^2 A \left(1 + \frac {\alpha_s (m_{W'})} \pi \right) \,, \\
  \Gamma(W'{}^+ \to t \bar{b}) &= 3 \kappa^2 A \left(1 + \frac {\alpha_s (m_{W'})} \pi \right) \left (1- \frac {m_t^2} {m_{W'}^2} \right)^2 \left(1+\frac 1 2 \, \frac {m_t^2} {m_{W'}^2}\right)\,,
  \label{eq:BRtb}
\end{align}
where $A = G_F m_W^2 m_{W'} / (6 \pi \sqrt{2})$ is an overall
constant~\cite{Rizzo:1981su,Rizzo:1981dm}. A first prediction of the
D-brane model is then that {\it $W'{}^+$ cannot be leptophobic}, with
comparable widths to dilepton and dijet channels.

The partial decay width to diboson is found to be
\begin{eqnarray}
  \Gamma(W'{}^+ \to W^+ Z) &= &\frac {A } 4 \lb\frac{g_L^2}{g_Y^2+g_L^2} \rb^2\, a^2 
\left( 1 - 2 \frac{m_W^2 + m_Z^2} {m_{W'}^2} + \frac {(m_W^2 - m_Z^2)^2} {m_{W'}^4} \right)^{3/2} \notag \\
& \times & \left(1 + 10 \frac{m_W^2 + m_Z^2} {m_{W'}^2} + \frac {m_W^4 + 10 m_W^2 m_Z^2 + m_Z^4} {m_{W'}^4} \right) \,, 
  \label{eq:BRWZ}
\end{eqnarray}
with $a = (\sin \phi/\xk) (m_{W'}^2 /
m_W^2)$~\cite{Brehmer:2015cia}. The second prediction of the D-brane
model is that (unless $\xk$ is small) {\it the decay rate into diboson
  is smaller than that into leptons:} \be \frac{\Gamma(W'{}^+ \to W^+
  Z)}{\Gamma(W'{}^+ \to ll)} \sim \frac{0.0121299 }{\kappa ^2}\, .
\ee Note, however, that $\xk$ is bounded from below $\xk
>0.55$~\cite{Brehmer:2015cia}. The reason is the following: the
unbroken $U(1)$ is associated with $T_L^3 + T_R^3 + B-L$ and therefore
(\ref{lambdayR}). As a result $g_R$ cannot be too small.

Following~\cite{Dobrescu:2015qna,Dobrescu:2015jvn}, we compute the
$W'$ production cross section for $pp$ collisions at $\sqrt{s} =
8$~TeV multiplying the leading-order cross sections computed with
MadGraph~\cite{Alwall:2014hca} (using model files generated with
FeynRules~\cite{Alloul:2013bka} and CTEQ6L parton
distributions~\cite{Pumplin:2002vw}, with factorization and
renormalization scales set at $m_{W'}$) by a scale-dependent
$K$-factor which takes into account next-to-leading order (NLO) QCD
effects.  We adopt $1.32 \alt K \alt 1.37$ as derived
in~\cite{Cao:2012ng}, which is larger than both the $K$-factor
obtained with the parton level Monte Carlo program for FeMtobarrn
processes (MCFM)~\cite{Campbell:2015qma} and the factor $K \approx
1.15$ computed in~\cite{Duffty:2012rf}. With this in mind, we
parametrize the $W'$ production cross
section at LHC8  by
\begin{equation}
\sigma(pp \to W') _{\sqrt{s} = 8~{\rm TeV}}  =816.686 \ g_R^2 \
\exp\left[ -3.53131 \left(\frac{m_{W'}}{{\rm TeV}}\right) \right]~{\rm pb} \, .
\label{sigma8}
\end{equation}
We note that, for $m_{W'} = 2~{\rm TeV}$, the value of the
parametrization is consistent with the results
of~\cite{Hisano:2015gna}. However, the results of our parametrization
are in general a factor of $\approx 8$ larger than the parametrization given in~\cite{Aguilar-Saavedra:2015iew}. We note that when we
multiply (\ref{sigma8}) by (\ref{eq:BRtb}) we recover the inclusive
cross section for $tb$ modes derived in~\cite{Duffty:2012rf}.

For $pp$ collisions at $\sqrt{s} = 13~{\rm TeV}$, we fit the results
derived in~\cite{Mitra:2016kov} for the $W'$ production cross section
at NLO with threshold resummation at next-to-next-to-leading logarithm
(NNLL) matched to threshold-improved parton distributions
functions. We parametrize the $W'$ production cross section at LHC13 by
\begin{equation}
\sigma (pp \to W')_{\sqrt{s} = 13~{\rm TeV}}  \simeq  3136.28 \
g_R^2  \ {\rm AntiLog}_{10} [ f(m_{W'})]~{\rm pb} \,,
\end{equation}
where
\begin{equation}
f(m_{W'}) = -1.61679  \left(\frac{m_{W'}}{{\rm TeV}}\right)
  +  0.10953  \left(\frac{m_{W'}}{{\rm TeV}}\right)^2  -  0.00385164  \left(\frac{m_{W'}}{{\rm TeV}}\right)^3  \, .
\end{equation}

\begin{figure}[tbp] 
\begin{minipage}[t]{0.44\textwidth}
    \postscript{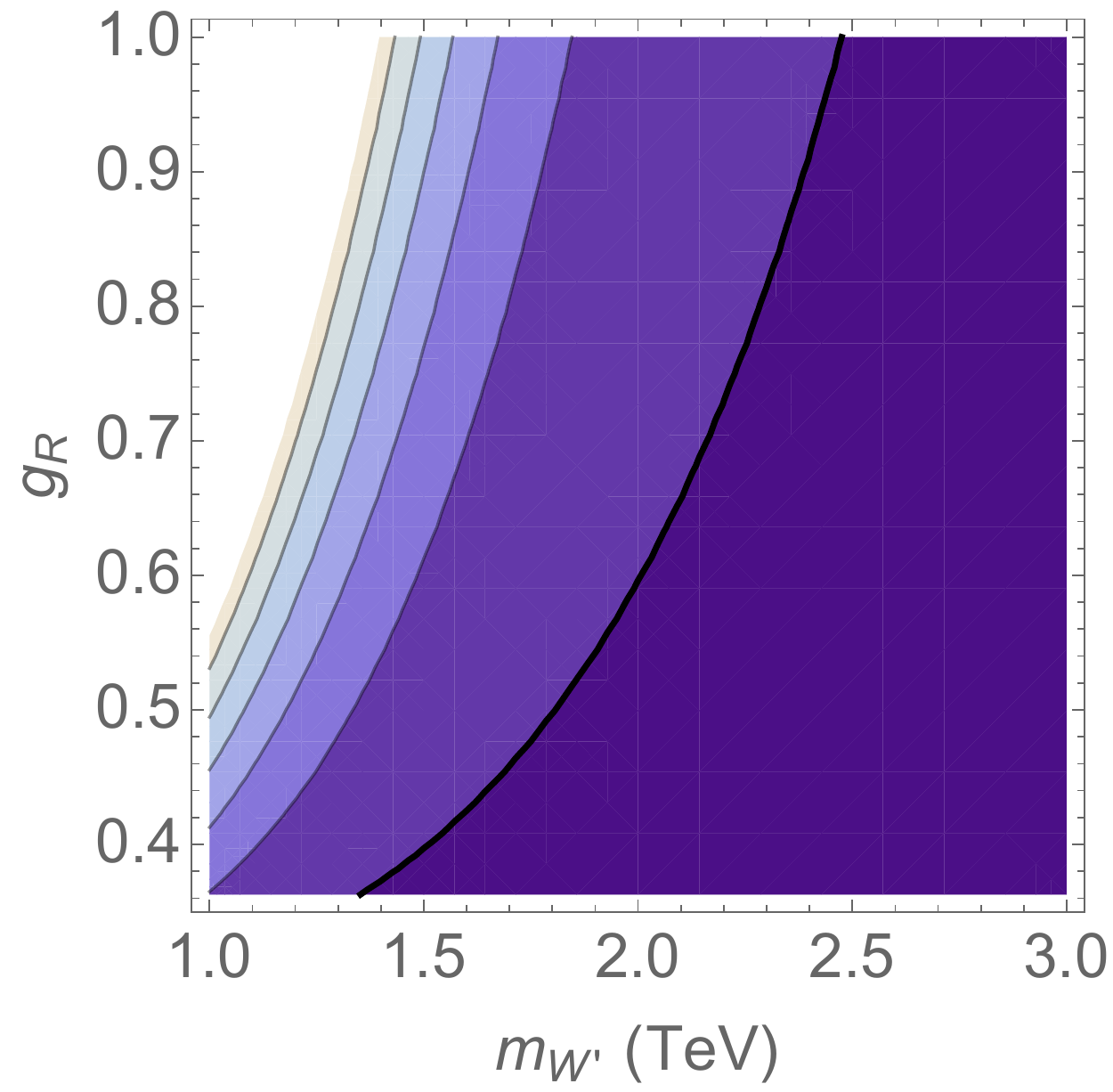}{0.99} 
\end{minipage} 
\begin{minipage}[t]{0.04\textwidth}
    \postscript{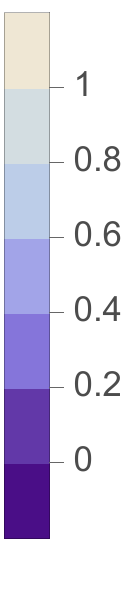}{0.99} 
\end{minipage}
\hfill \begin{minipage}[t]{0.44\textwidth}
  \postscript{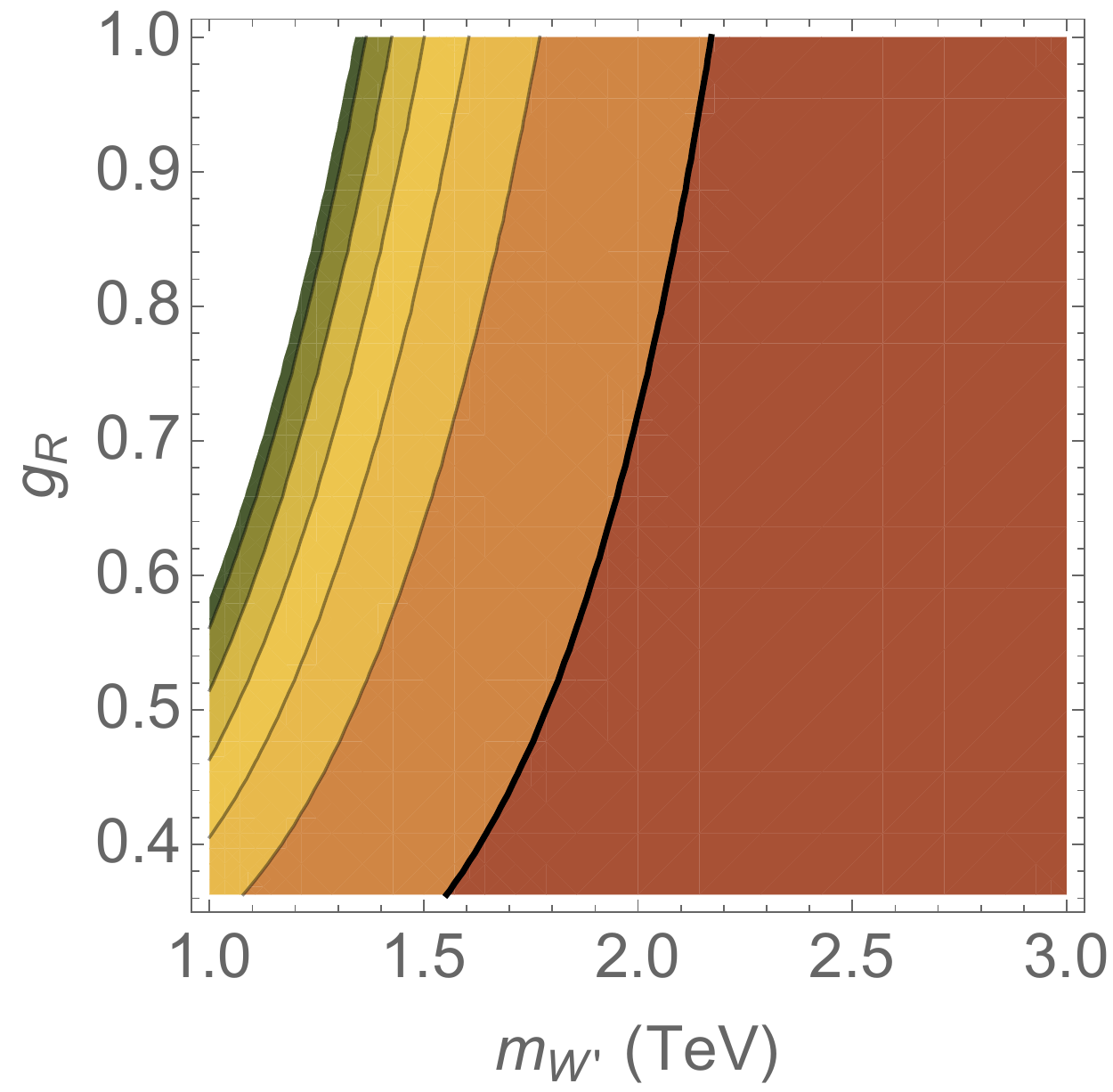}{0.99} 
\end{minipage} 
\begin{minipage}[t]{0.04\textwidth}
  \postscript{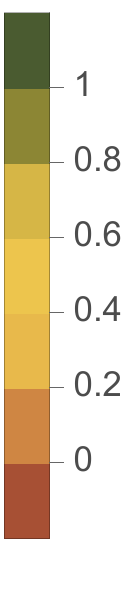}{0.99} 
\end{minipage} 
\caption{Contours of constant $\Delta \sigma_X$ for $X =$ dijet (left)
and $tb$ (right) modes for a  generous range of $g_R$, and LHC8 collisions. }  
\label{LHC8} 
\end{figure}
\begin{figure}[tbp] 
\begin{minipage}[t]{0.44\textwidth}
    \postscript{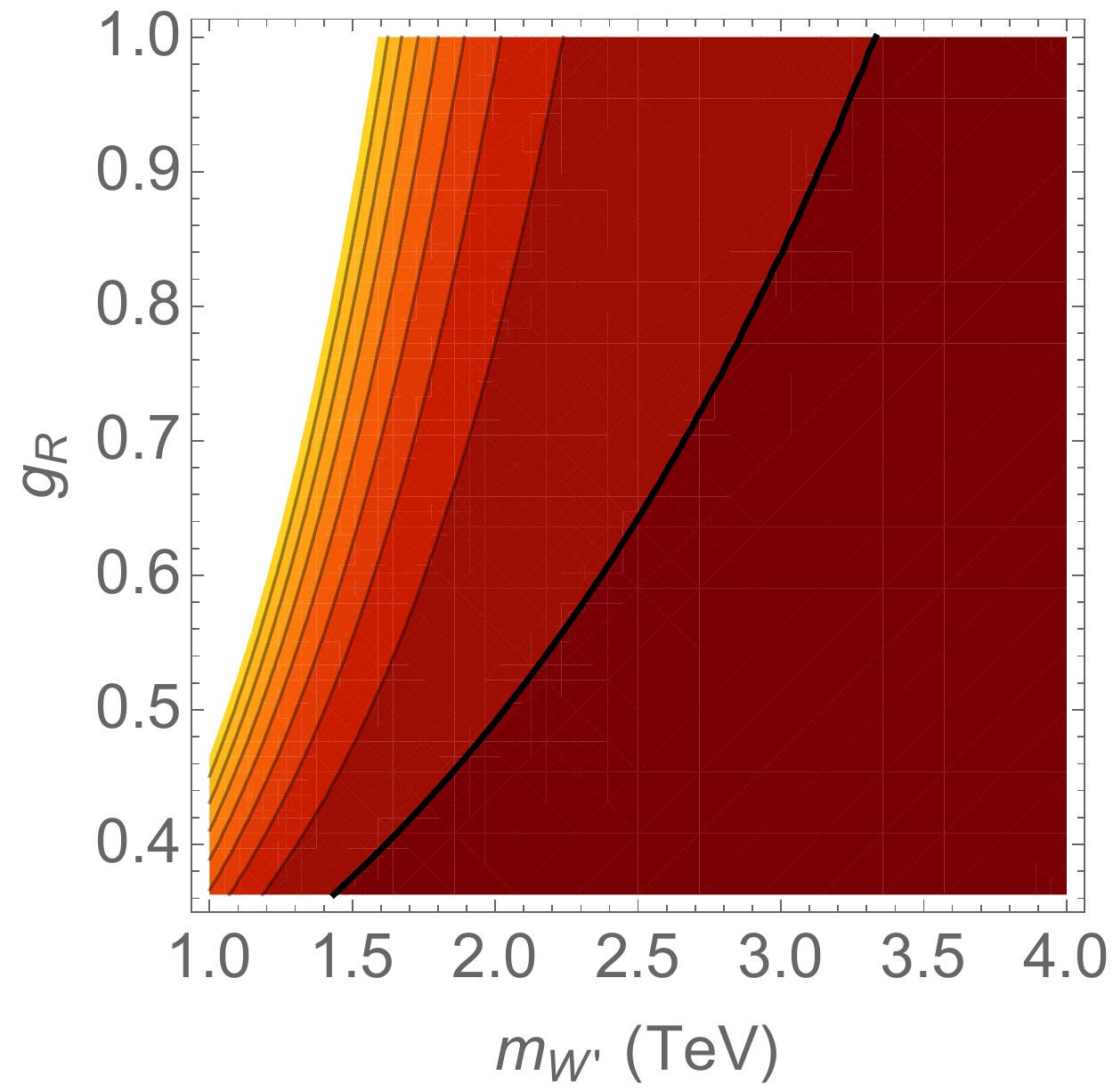}{0.99} 
\end{minipage} 
\begin{minipage}[t]{0.04\textwidth}
    \postscript{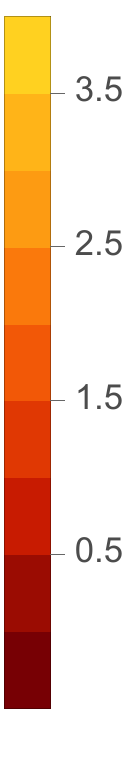}{0.99} 
\end{minipage}
\hfill \begin{minipage}[t]{0.44\textwidth}
  \postscript{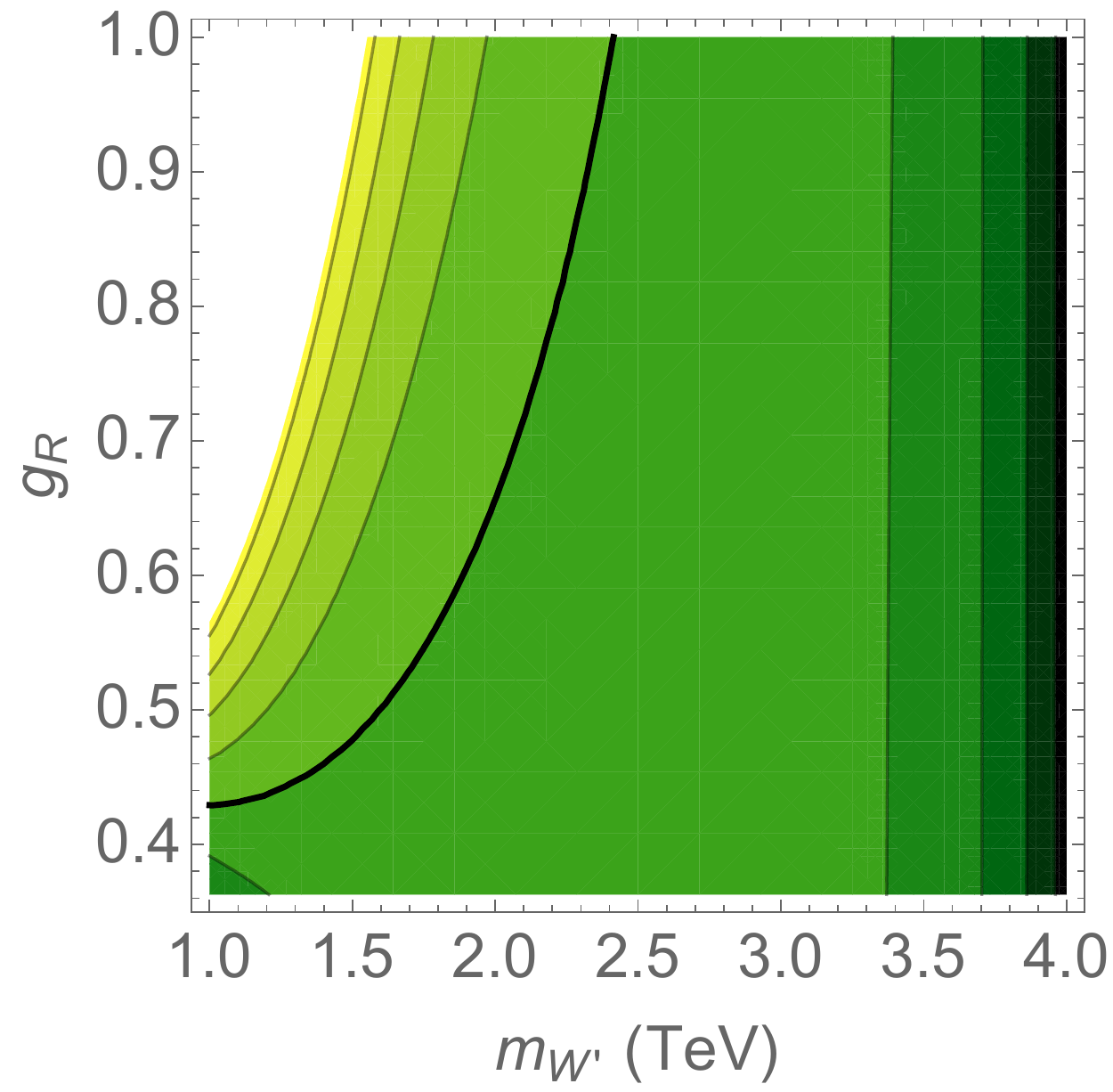}{0.99} 
\end{minipage} 
\begin{minipage}[t]{0.04\textwidth}
  \postscript{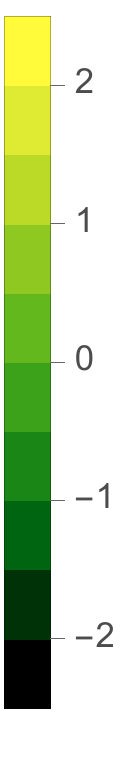}{0.99} 
\end{minipage} 
\caption{Contours of constant $\Delta \sigma_X$ for $X =$ dijet  (left)
and $tb$ (right) modes for a  generous range of $g_R$, and LHC13 collisions.}  
\label{LHC13} 
\end{figure}

We now turn to bound the parameter space of the D-brane model. To do so, we define 
\begin{equation}
\Delta \sigma_X = \sigma (pp \to W') \times {\cal B} (W' \to X) -
\left[\sigma (pp \to W') \times {\cal B} (W' \to X) \right]_{95\%{\rm
    CL}} \,,
\end{equation}
where $\left[\sigma (pp \to W') \times {\cal B} (W' \to X) \right]_{95\%{\rm
    CL}}$ is the 95\%~C.L. upper limit on the inclusive cross
section $\sigma (pp \to W') \times {\cal B} (W' \to X)$, with $X =$
dijet and $tb$ modes.  Our results are encapsulated in
Figs.~\ref{LHC8} and \ref{LHC13} where we show contours of constant $\Delta
\sigma_X$ for $pp$ collisions at $\sqrt{s} = 8~{\rm TeV}$ and $\sqrt{s} =
13~{\rm TeV}$.  The dark line corresponds to $\Delta \sigma_X =0$,
with the parameter space to the left being excluded at the 95\%
C.L. In the plot we also show contours to the left of the exclusion
line in order to leave room to move the bound as necessary to account
for detector effects and possible future change in theoretical
uncertainties. Note that the exclusion line is conservative since our
calculation is at the parton level. Therefore, taking into account the
aforementioned effects would move the line to lower masses. As can be
seen in Figs.~\ref{LHC8} and \ref{LHC13},  for the interesting range $0.48 <
g_R({m_W'}) <0.60$, gauge bosons with $m_{W'} \agt 2.5~{\rm TeV}$ are
not excluded. Note that at present the most restrictive bounds on $W'$ masses are
from dijet searches, but an actual discovery will perhaps be possible using
the $tb$ modes, which have comparable sensitivity.

\section{Conclusions}
\label{s4}

In this paper we studied extensions of the SM in which restoration of the
left-right symmetry is pulled down to the TeV-scale ballpark.  We considered string
theory setups in which the gauge theories live on D($3+p$)-branes
which entirely fill the uncompactified part of space-time and wrap
certain $p$-cycles inside the compact six-dimensional manifold. The
chiral matter fields appear at the intersection of two
D$(3+p)$-branes. To develop our program in the simplest way, we worked
within the construct of a minimal model characterized by $U(3)_C \otimes Sp(1)_L
\otimes U(1)_L \otimes Sp(1)_R$.

The approach we have taken in this work can be regarded as an
effective theory with both interesting LHC phenomenology and unique
theoretical characteristics such as conservation of $B$ to
prevent proton decay to all orders in perturbation theory and
violation of $L$ without Majorana masses.  The absence of
Majorana masses constrains the $W'$ decay rates, yielding comparable
widths to dilepton and dijet channels. This naturally narrows the
parameter space for LHC searches while providing at the same time a
definite prediction of D-brane models. Additional abelian gauge
symmetries, inherent to the structure of D-brane gauge theories,  
can provide interesting corroboration for string physics near the TeV-scale.

We have derived bounds on the $(g_R, m_{W'})$ plane using the most
recent searches at LHC13 Run II with 2016 data. Our analysis indicates that
independent of the gauge coupling $g_R$, right-handed $W'$ bosons with masses above
about 3.5~TeV are not ruled out. As the LHC is just beginning to probe the
TeV-scale, significant room for discovery remains.

\acknowledgments{The research of L.A.A.  is supported by U.S. National
  Science Foundation (NSF) Grant No.~PHY-1620661 and by the National
  Aeronautics and Space Administration (NASA) Grant No. NNX13AH52G.
  D.L. is partially supported by the ERC Advanced Grant Strings and
  Gravity (Grant No. 32004) and by the DFG cluster of excellence
  ``Origin and Structure of the Universe.''  T.R.T. is supported by
  NSF Grant No. PHY-1620575.  He is grateful to the United States
  Department of State Bureau of Educational and Cultural Affairs
  Fulbright Scholar Program and to Polish-U.S. Fulbright Commission
  for a Fulbright Award to Poland. Any opinions, findings, and conclusions or
  recommendations expressed in this material are those of the authors
  and do not necessarily reflect the views of the National Science
  Foundation.}

\end{document}